\documentclass[aps,twocolumn,showlabels,showrefs,amsmath,amssymb,pre,superscriptaddress, floatfix, colors]{revtex4-1}

\usepackage{amssymb,amsmath,amsfonts}
\usepackage{graphicx,color}
\usepackage[english]{babel}
\usepackage{tabularx}

\usepackage{soul}

\newcommand{\Sum}[2]{{\sum\limits_{#1}^{#2}}}

\begin{document}

\title{Unraveling Modular Microswimmers: From Self-Assembly to Ion-Exchange Driven Motors}

\author{Benno Liebchen}
\email{liebchen@hhu.de}
\affiliation{Institut f\"{u}r Theoretische Physik II: Weiche Materie, Heinrich-Heine-Universit\"{a}t D\"{u}sseldorf, D-40225 D\"{u}sseldorf, Germany}
\author{Ran Niu}
\email{ranniu@uni-mainz.de}
\affiliation{Institut f\"ur Physik, Johannes Gutenberg-Universit\"at Mainz, Staudingerweg 7, D-55128 Mainz, Germany}
\author{Thomas Palberg}
\affiliation{Institut f\"ur Physik, Johannes Gutenberg-Universit\"at Mainz, Staudingerweg 7, D-55128 Mainz, Germany}
\author{Hartmut L\"owen}
\affiliation{Institut f\"{u}r Theoretische Physik II: Weiche Materie, Heinrich-Heine-Universit\"{a}t D\"{u}sseldorf, D-40225 D\"{u}sseldorf, Germany}

\date{\today}

\begin{abstract}
Active systems contain self-propelled particles and can spontaneously self-organize into 
patterns making them attractive candidates for the self-assembly of smart soft materials. 
One key limitation of our present understanding of these materials hinges on the complexity 
of the microscopic mechanisms driving its components forward. Here, by combining experiments, 
analytical theory and simulations we explore such a mechanism
for a class of active system, modular microswimmers, 
which self-assemble from colloids and ion-exchange resins on charged substrates.
Our results unveil the self-assembly processes and the working mechanism of the
ion-exchange driven motors underlying modular microswimmers, 
which have so far been illusive, even qualitatively. 
We apply these motors to show that modular microswimmers can
circumvent corners in complex environments and move uphill. 
Our work closes a central knowledge gap in modular microswimmers and 
provides a facile route to extract mechanical energy from ion-exchange processes.
\end{abstract}

\maketitle

\section{Introduction}
Self-propulsion of biological agents like bacteria, crawling cells or actin-based gels is 
involved in most processes in human and animal life, from its beginning during embryogenesis 
to the motion of muscles and the emergence of diseases like Alzheimer or cancer metastasis 
formation sometimes ending life. While most self-propelled biological agents \cite{1} 
are too complex to directly understand them microscopically, the past decade has led to 
the development of synthetic microswimmers \cite{2,3,4,5,6}.
 Praised for the simplicity of their design, synthetic swimmers present a promising platform to develop and 
reproducibly test an understanding of the properties, designability and functionality of "active" materials containing 
self-propelled particles \cite{7,8,9}.
\begin{figure}
\includegraphics[width=0.48\textwidth]{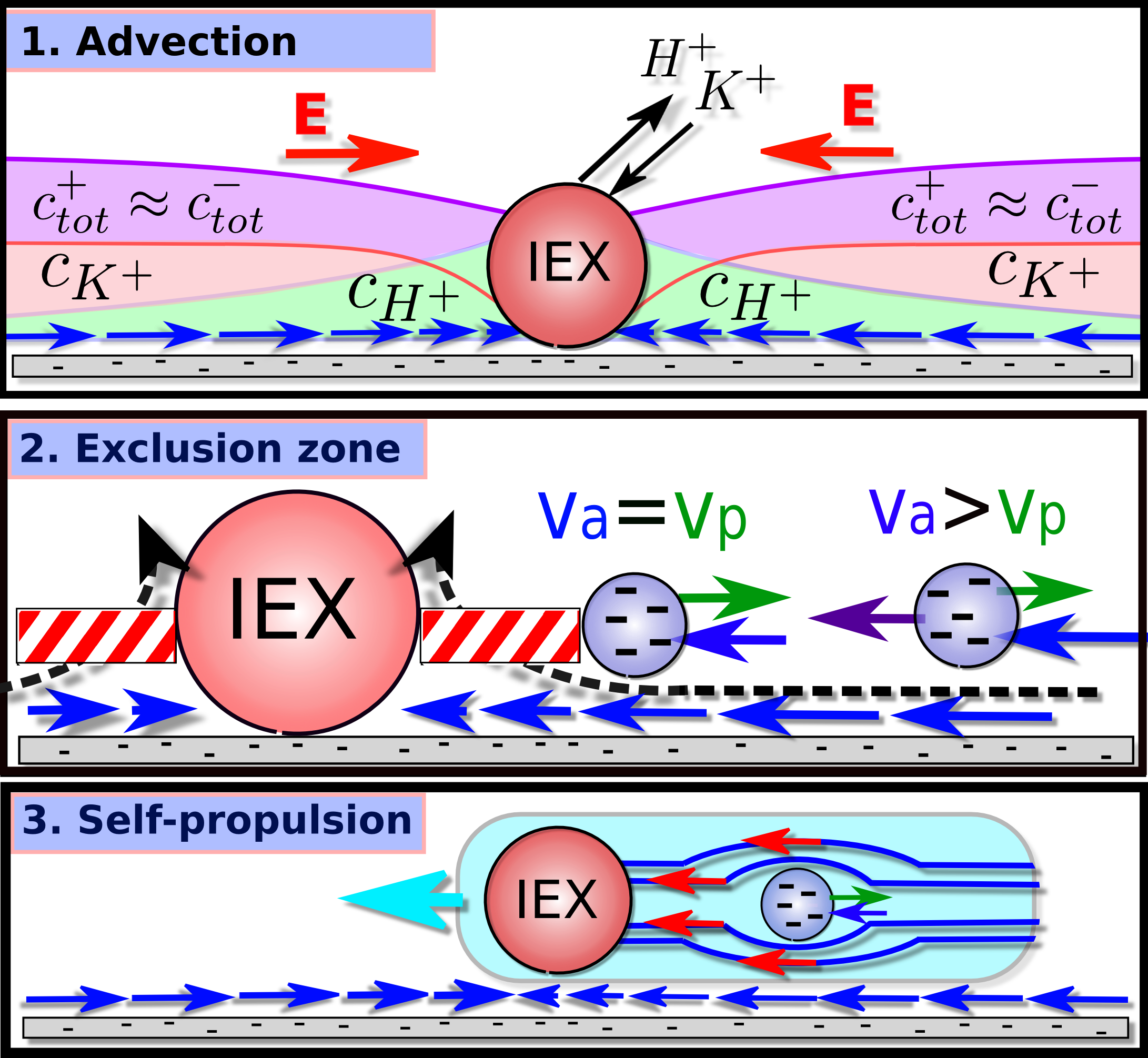}
\caption{Self-assembly and modular swimming mechanism (side view). 1. \textbf{Advection}: The resin (IEX) exchanges $K^+$-ions against more mobile 
$H^+$-ions leading to long-range ionic gradients evoking a spontaneous electric field \textbf{E}. 
This field acts on the charged fluid inside the substrate double layer inducing fluid flow 
towards the resin (bottom arrows, blue). 
2. \textbf{Exclusion zone}: At large distances colloids (spheres with $-$ symbols) 
advect towards the resin, but slower than the surrounding solvent due to diffusiophoresis 
in the charge-neutral ionic gradients created by the resin ($v_a>v_p$). 
Close to the resin, incompressibility obliges the solvent to stream upwards (dashed arrow), 
reducing advection along the substrate. Thus, phoresis dominates at short distances so that the colloid settles at a distance 
where advection and phoresis balance ($v_a=v_p$) defining an "exclusion zone" (hatshed rectangle). 
3. \textbf{Self-propulsion}: The action of the chemical gradients on the counterions in the 
Debye layer of the charged and stalled colloid creates an osmotic flow towards the resin. The resin sees an enhanced solvent 
flow coming from the direction of the colloid leading to its advection.
}
\label{fig1}
\end{figure}
It has turned out, however, that despite their minimalistic design, the most basic question - how synthetic 
microswimmers precisely move and interact - is remarkably complex to answer for most examples. In fact, their 
self-propulsion typically involves coexisting gradients in different phoretic fields (neutral chemicals, ions, temperature fields) 
coupling to the solvent and contributing to swimming by different, sometimes competing, phoretic mechanisms \cite{10,11,12}. 
However, besides generating unwanted complexity in the swimming mechanism, phoretic fields also cross-couple 
different microswimmers inducing a remarkably versatile collective behaviour 
including clusters, travelling waves and rotating gears, releasing a huge potential for 
nonequilibrium self-assembly \cite{13,14,15,16,17,18,19,20,21,22,23,24,25,26,99}. 
Like the swimming mechanism itself, our current understanding of the collective behaviour of synthetic microswimmers 
\cite{13,14,15,16,17,18,19,24,25,26,27,28,29,103} commonly suffers from a lack of knowledge of the relevant fields and the 
corresponding coupling coefficients \cite{23} (which is an issue also in other fields \cite{101,102}).
Here, we unravel (microscopic) mechanisms which dictate the dynamic self-assembly \cite{30} and 
propulsion mechanism of an intensively studied class of synthetic microswimmers: 
modular swimmers \cite{31,32,33,34}. The here studied species self-assembles from 
ion-exchange resins (IEX) and passive colloids. It spontaneously starts moving when 
both components bind, typically leaving a pronounced exclusion zone in between. 
While observed in various experiments \cite{31}, only the attraction between resin 
and colloids has been explained \cite{35,36,100}. 
However, \emph{both the formation of an exclusion 
zone and the very fact that the modules can self-propel remains elusive, even qualitatively.}
(Note that ref.~\cite{100} has discussed the mechanism underlying self-propulsion of the swimmers and has 
associated it with a flow from colloid to resin, but has neither fully explained the
origin of this flow, nor the mechanism leading to the exclusion zone.) 
The key advancement of the \emph{present work is to provide 
a physical picture for the whole self-assembly 
and swimming mechanism of modular swimmers} (Fig.~\ref{fig1}).
We summarize the key aspects of 
this picture in a (microscopically justified) minimal model whose predictions are in close 
quantitative agreement with our experiments, both on flat and on tilted substrates. 
In the latter case, we report, for the first time, (positive) gravitaxis in modular swimmers, 
which we apply to guide them in microgroove-imprinted substrates around corners and uphill. 
The latter shows that modular swimmers can serve as motors extracting mechanical energy from 
ion-exchange processes.
\section{Setup}
We consider a cylindrical cell with top and bottom confining glass plates containing a cationic IEX (radius $\sim 22.5\mu m$) 
and monodisperse polystyrene colloids (radii $\sim 7.6 \mu m$) in deionized, degassed water. (Downsizing of both IEX and colloids is straightforward, at least 
to scales $\sim 1\mu m$.)
Several ions may contribute to the electrolyte: $Na^+$ or $K^+$ are released from the substrate; 
likely negative ions are $Cl^-$, $HCO_3^-$ coming from air, and $OH^-$ ions in low concentration (from autoprotolysis $ 0.1 \mu mol/L$). 
$H^+$ ions in turn emerge from the resin surface in exchange for $K^+$ ions. 
For our theory, we exploit that $Cl^-$, $Na^+$ or $K^+$ and $HCO_3^-$ all have roughly similar diffusion coefficients 
$D \sim (1.2-2)\times 10^3\mu m^2/s$ and account for only three (effective) species: $K^+$, $Cl^-$ and $H^+$; see also \cite{35}.
\section{Experiments}
\begin{figure*}[htb]
\includegraphics[width=0.95\textwidth]{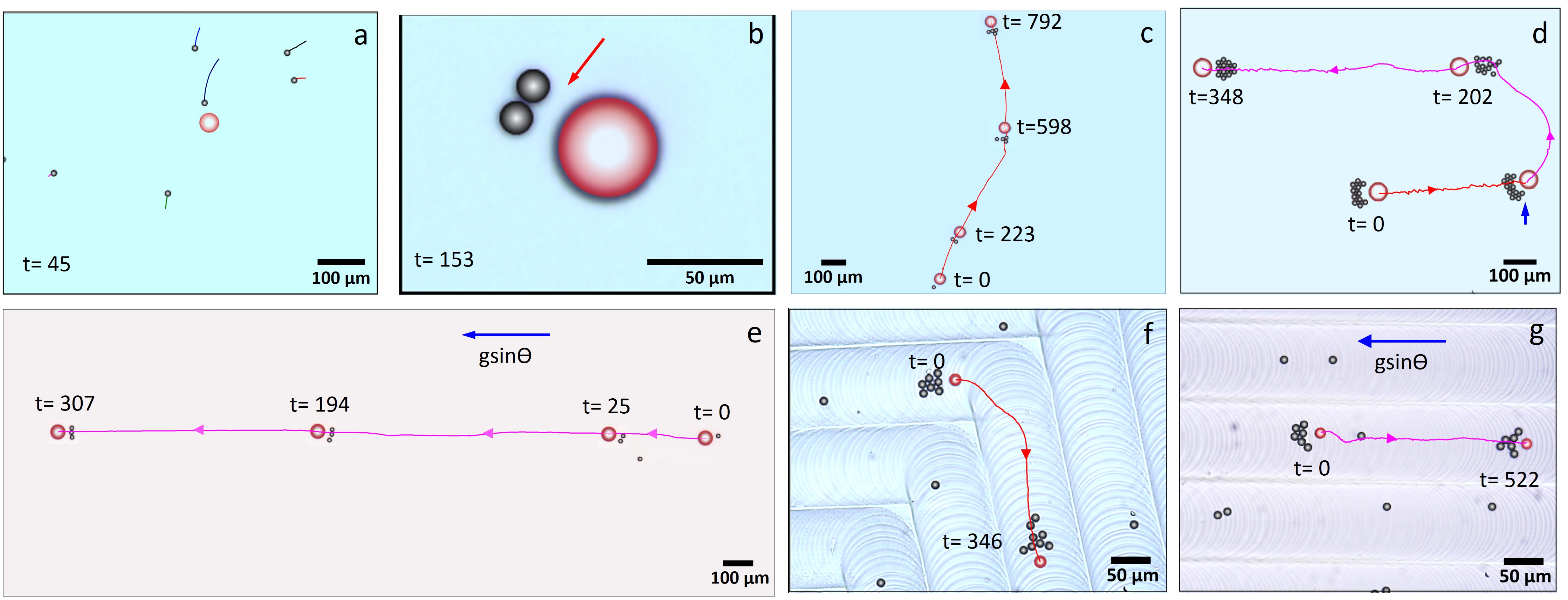}
\caption{Experiments: a) colloids moving towards the IEX along a glass plate. b) When approaching the resin, the colloids leave an exclusion zone (arrow) to the surface of the resin and form a self-propelling modular-swimmer. c) Trajectory of a modular swimmer collecting colloids along its path. d) Downhill swimming: gravity induces downhill swimming (gravitaxis). e) 
Self-assembly of a modular swimmer on a tilted substrate. f-g) Microgrooves guide modular swimmers around corners (f) and uphill (g).} 
\label{fig2}
\end{figure*}

Colloids move towards the IEX (Fig.~\ref{fig2}a), self-aggregate there and bind to the IEX, 
leaving a significant exclusion zone of $\gtrsim$10$\mu m$ surface-to-surface distance (Fig.~\ref{fig2}b, video 1).
 Once the first colloid binds to the IEX, the complex starts to move autonomously (Fig.~\ref{fig2}b) 
and picks up new colloids binding to the back of the resin (Fig.~\ref{fig2}c, video 2). 
Here, the speed of the complex increases; when about $5-10$ colloids have bound to the IEX, the modular swimmer, 
quite abruptly, reaches a plateau speed (Fig.~\ref{fig4}) and no further colloids can permanently bind to the resin.
Note that the speed of the complex fluctuates in time by about $20\%$ of its mean velocity, 
partly due to dynamic shape changes of the colloidal tail.
Tilting the substrate, as expected, a single IEX slides or rolls downwards (not shown). When colloids bind to it, as before, 
the complex starts self-propelling in directions pointing away from the colloids. 
Ultimately, however, it always turns downhill (Fig.~\ref{fig2}d,e, video 3), i.e. modular swimmers show positive gravitaxis. 
We now prepare the setup on a substrate with imprinted arc-shaped microgrooves (5$\mu m$ depth, $90\mu m$ width) 
(see Supporting Information, SI for further details) 
and exploit gravitaxis forcing the swimmers to move along the groove bottom. 
Thus, curved grooves allow guiding the swimmers around corners (Fig.~\ref{fig2}f) and grooves in tilted substrates can be used to induce persistent uphill swimming (Fig.~\ref{fig2}g). 
The latter can be used, in principle, as a mechanism to extract mechanical work from ion-exchange processes, i.e. to store energy. 
The energy storage rate is about $\sim 0.1 pJ/h$,
allowing to store several $pJ$ per modular swimmer, since the ion-exchange resin depletes on timescale of one to several days.
While the aggregation of colloids towards the resin is broadly understood \cite{35}, we now explore the self-assembly and self-propulsion mechanism of modular microswimmers.
\section{Solvent advection towards the resin}
The IEX creates a solvent flow along the glass plate advecting colloids towards it by the following mechanism (Fig.~\ref{fig1} top panel): The 
IEX exchanges $K^+$-ions with $H^+$-ions creating a dip in the density of $K^+$-ions close to the resin surface and a surplus of $H^+$-ions as monitored by micro-photometry \cite{37}. Since the latter species is more mobile, it diffuses rapidly away from the resin provoking a dip in the overall density of positive charges close to the resin surface dying out (algebraically) slowly with distance to the resin (see Fig.~\ref{fig1}, top panel and SI for details). This tendency for a charge imbalance creates a spontaneous (unscreened ~\cite{38}) electric field inducing motion of both positive and negative ions in a way preventing a charge-imbalance; the steady-state result is a long-range-inhomogeneous density profile which is almost identical for positive and negative ions (This remains true in the presence of solvent advection; proof in SI).
The unscreened electric field drags positive charges in the Debye layer of the negatively charged substrate towards the resin; 
this generates a stress in the solvent leading to flow along the substrate, towards the resin. 
This flow is complemented by (charge neutral) diffusioosmosis hinging on steric and dipolar interactions of 
the nonuniformly distributed ions (counter-oriented gradients of different cation species) with the substrate surface. 
The resulting flow advects colloids towards the resin. Quantitatively, a solvent with viscosity $\eta$  and electrical permittivity $\varepsilon$ advects parallel to the bottom glass plate 
towards the resin with a velocity (see SI for details):
\begin{equation}
{\bf v}_a(\rho)\approx \frac{-\epsilon k T}{4\pi \eta e} \left[\zeta \mathcal{D} \frac{\partial_\rho c_{H^+}}{c^0_{K^+}} - 2\frac{k T}{e}\ln(1-\gamma^2) \frac{\partial_\rho c_{Cl^-}}{c_{Cl^-}^0}\right]{\bf e}_\rho
\label{phoretic_velo}
\end{equation}
Here, we have applied the classic expression for the phoretic motion of colloids due to diffusiophoresis in a charge neutral ionic solvent \cite{39}
and the following notion: ${\bf e}_\rho$ is the unit vector pointing radially away from the resin along the substrate and
$c_{H^+}\equiv c_{H^+}(\rho);\;\; c_{Cl^-}\equiv c_{Cl^-}(\rho)$
are the ionic concentration profiles (number densities)
of $H^+$- and $Cl^-$-ions at a distance $\rho$ from the resin surface and denotes the derivative with respect to $\rho$.
$c_{K^+}^0$, $c_{Cl^-}^0$ are the background densities of $K^+$- and $Cl^-$-ions far away from the resin,
$k T$ is the thermal energy, $e$ is the elementary charge, $\zeta$ is the zeta potential of the substrate (glass plate), 
$\mathcal{D}=(D_{H^+}-D_{K^+})/(2D_{K^+})$ is a reduced
diffusion coefficient of order 1 and
$\gamma=\tanh(\xi)$ with $\xi=\zeta e/(4  k T)$.
\\Fluid motion is driven here by diffusioosmosis in charge-neutral ionic concentration gradients which has two components: 
(i) electroosmosis based on the motion of charged fluid elements within the double layer of the substrate driven by the spontaneous electric field and (ii) 
neutral diffusioosmosis (also called chemiosmosis) which is based on interactions of the solute particles with the 
colloid surface and 
arises from a tendency of the system to 
reduce the surface free energy. This component does not involve electric fields; compare \cite{39} regarding the terminology used here. 
Both components are of similar strength here (SI). 
By calculating the relevant concentration gradients, we find both for moderate and large distances a $1/\rho$-far field scaling,
if the cell height $L$ is not too large. 
For large $L$, we predict a $1/\rho^2$-far field scaling (SI). 
Both agrees with previous experimental findings \cite{35}. We also provide a rough ab initio prediction of the near-field saturation value for the solvent advection speed of about $3.7 \mu m/s$ for the present experimental conditions (SI). This is also consistent with experiments \cite{35}.
\section{Exclusion zone formation}
Like the bottom glass plate, the colloids are negatively charged. Hence, the spontaneous electric field creates a stress in the interfacial layer of the colloids inducing a surface slip. 
This, in turn, creates a phoretic motion of the colloids pointing away from the resin and opposite to the solvent flow. At large distances to the resin, 
this phoretic motion is about 3 times weaker than advection so that colloids move towards the resin (SI), 
but roughly $1/3$rd
slower than the surrounding solvent. Conversely, close to the resin, solvent incompressibility obliges the solvent to move upwards (Fig.~\ref{fig1}, middle panel). This reduces colloidal advection along the substrate but does not affect their phoretic motion. As a consequence, advection towards the resin dominates at large distances to the resin whereas phoretic motion away from it dominates at short distances. This leads to a stable equilibrium configuration between colloids and resin (exclusion zone).
\\Quantitatively, the same expression (1) describing the slip velocity over the bottom substrate also describes the slip velocity over the colloidal surface with the zeta potential being replaced. The colloidal phoretic velocity relative to the solvent reads ${\bf v}_p=-\langle {\bf v}_s({\bf r}) \rangle$ with brackets denoting the surface average \cite{39} and ${\bf v}_s({\bf r})$ being the surface slip velocity which we evaluate at the colloid-midpoint for simplicity (compare \cite{40} for more rigorous arguments). The overall velocity of a colloid relative to the fixed substrate thus reads ${\bf v}_c={\bf v}_a+{\bf v}_p$ where ${\bf v}_p$ points radially away from the resin.
\section{Self-propulsion mechanism}
Experimentally, once a colloid approaches the resin, the complex starts to move. Here, we propose the following picture: as for the substrate, the action of the long-ranged chemical gradients on the counterions in the Debye layer of a charged and stalled colloid creates a surface slip. Since the colloid is stalled, this surface slip is essentially an osmotic flow pointing towards the resin (Fig.~\ref{fig1}.3). Thus, the resin sees an enhanced solvent flow coming from the direction of the colloid leading to its advection. Quantitatively, we model this additional solvent velocity (outside the colloidal double layer) as
\begin{equation}
  {\bf u}_b({\bf r})\sim \frac{-1}{2}\left(\frac{R}{r}\right)^3 \left(\frac{3{\bf r}{\bf r}}{r^2}-I\right)\cdot {\bf v}_c(r) \label{bfv}
\end{equation}
where we understand ${\bf r}$ as the (shortest) surface-to-surface distance vector from the resin to a colloid with radius $R$.
Note that fixing the colloid with a force in bulk would induce a Stokeslet; here, since stalling occurs due to phoretic motion we phenomenologically use the (negative) velocity-field created by force-free phoretic motion instead \cite{41}, but note that the actual flow field is probably more complicated e.g. due to substrate influences. Being significantly faster than the resin, colloids follow the moving resin at an almost constant distance. When several colloids bind to the resin, they collectively push the resin forward. Overall, since additional colloids bind at larger distance to the resin, the complex speed enhances sublinearly with the number of colloids involved, as in experiments (Fig.~\ref{fig4}).
\\Crucially, at some point, the resin speed becomes comparable to the colloid advection speed. Then colloids bind only briefly to the resin enhancing the resin speed over the critical value and disconnect. This is precisely as in experiments.
\begin{figure}
\includegraphics[width=0.48\textwidth]{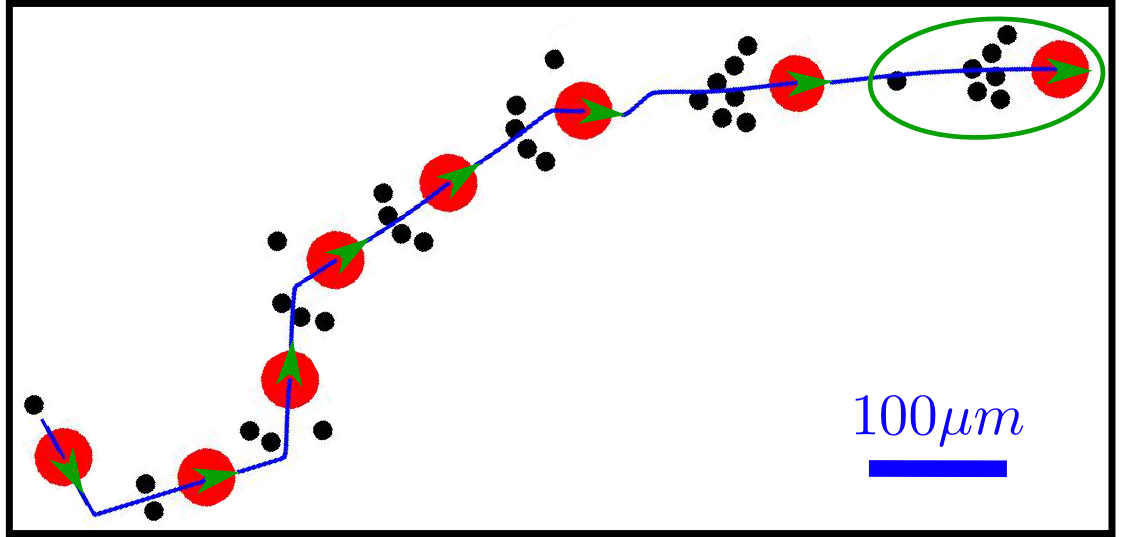}
\caption{Simulated self-assembly dynamics of a modular swimmer based on 
Eqs.~(\ref{model1}--\ref{model5}). As in our experiments, colloids (small spheres, black) 
advect towards the IEX (large spheres, red), bind to it leaving an exclusion zone to the resin surface. 
The complex then starts self-propelling; further colloids bind to the resin, forming rows 
at the back of the resin, until at some point, colloids can no longer follow the moving resin 
(ellipse, green). Model parameters $b_c=20\mu m^2/s; b_s=100\mu m^2/s$; $R_{IEX}=22.5\mu m,R_c=7.6\mu m$.}
\label{fig3}
\end{figure}
\begin{figure}
\includegraphics[width=0.48\textwidth]{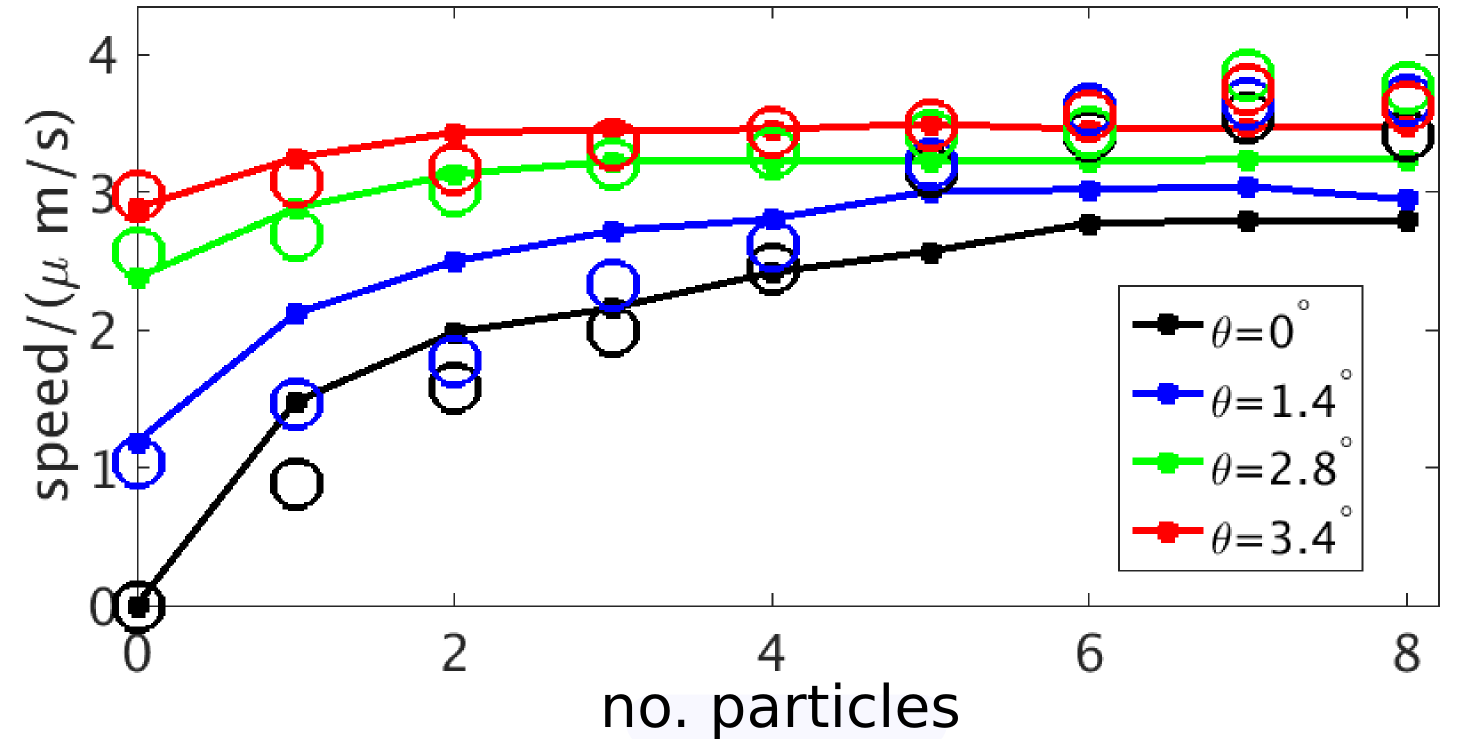}
\caption{Swim speed in experiments (circles) 
and simulations (squares with lines) without (bottom symbols and line; black) and with gravity (other symbols and lines, colors);  
substrate tilt angle given in the key. Experimental values are averages over 60-90 complexes (standard deviation up to $0.2\mu m/s$, indicated by circle sizes). 
Parameters as in Fig.~\ref{fig3} and 
$\tilde m_{IEX}=18.7$ (fitted for zero colloids attached) and $\tilde m_c=0.097$ 
(estimated based on mass and size ratio of colloids and IEX).}
\label{fig4}
\end{figure}
\section{Model}
We now summarize the described physical picture and the discussed equations in a minimal model. 
At large distances to the resin $\rho \gg \nu$, the solvent speed reads $v_a=b_s/\rho$ (SI), where 
$\nu$ sets the length scale at which the direction of solvent motion starts pointing upwards due to solvent incompressibility; its value can be fixed by comparing the size of the 
exclusion zone with experiments. 
Thus, we model the advective speed as
\begin{equation}
{\bf v}_a=-\frac{b_s}{\rho} \tanh[(\rho/\nu)^2] {\bf e}_\rho \label{model1}
\end{equation}
where $\rho$ is the 2D distance from resin surface to colloid surface, ${\bf e}_\rho$ points radially away from the resin along the substrate
and the $\tanh$ has been chosen phenomenologically, as a convenient choice to interpolate between near field and far-field behaviour.
As discussed, the phoretic speed of a colloid follows the same law as the solvent-advection but with opposite sign and an about $3-4$ times smaller coefficient (SI). Since the phoretic motion of a colloid is basically unaffected by the direction of solvent flow it reads
\begin{equation}
  {\bf v}_p=\frac{b_c}{\rho}{\bf e}_\rho \label{model2}
\end{equation}
pointing radially away from the resin. The competition between Eqs.~(\ref{model1}) and (\ref{model2}) determines the exclusion zone. Finally, following Eq.~(\ref{bfv}) we model the resin speed as
\begin{equation}
  {\bf v}_{b} = -\Sum{j=1}{N}\left(\frac{R_c}{\rho_j}\right)^3 {\bf v}^j_p \label{model3}
\end{equation}
decaying with a $1/\rho^4$-law from the $j$-th colloid ($j=1,..,N$) with speeds ${\bf v}^j_p$ given by Eq.~(\ref{model2}) and distances
$\rho_j$ measured to the resin-surface. Finally, we model short-range steric repulsions among the colloids with a purely repulsive Lennard-Jones potential $U$ (truncated and shifted). Overall, we have
\begin{equation}
  \dot {\bf r}_{IEX} =  {\bf v}_b + \frac{1}{\gamma_{IEX}}{\bf F}_{IEX}, \label{model4}
\end{equation}
\begin{equation}
  \dot {\bf r}_C^{i} =  {\bf v}^{i}_a + {\bf v}^{i}_p+ \frac{1}{\gamma_C}{\bf F}_{C} + \frac{1}{\gamma_C}\nabla_{{\bf r}_i} U \label{model5}
\end{equation}
where $\gamma_{IEX,C}=6\pi R_{IEX,C} \eta$ is the friction coefficient of resin and colloids, respectively.
Here, we have simply added gravitational forces ${\bf F}_{IEX},{\bf F}_C$ to Eqs.~(\ref{model4},\ref{model5}),
occurring for tilted substrates. These forces read
${\bf F}_{IEX,C} = g \tilde m_{IEX,C} \tan\theta$ where
$\tilde m_{IEX,C}$ is the effective mass of the resin and colloids respectively accounting for the facts that the particles roll (and slide) downhill, relative to the surrounding solvent. 
(Note here that gravity causes additional $1/\rho$-flow far fields, which we neglect here since their coefficients are very small for the considered tilt angles.)
Since the resin density exceeds that of the colloids, the swimmer is bottom-heavy, leading to positive gravitaxis.
We now numerically solve Eqs.~(\ref{model4},\ref{model5}) using periodic boundary conditions and random initial conditions for colloids and resin. 
At early times we observe colloids moving towards the resin; they stop $10-15 \mu m$ before the resin surface creating an exclusion zone (Fig.~\ref{fig3}), as in experiments. 
The complex then starts to self-propel and subsequently attracts more colloids which self-assemble in rows behind the resin (Fig.~\ref{fig3}), all as in experiments (Fig.~\ref{fig2}c). 
Also as in experiments, at some point, the complex reaches a speed comparable to the advection speed of the colloids; then 
no further colloids can bind to the resin and stripe off (green oval in Fig.~\ref{fig3}).
\\We now compare the swimming speed more quantitatively: Fig.~\ref{fig4} shows that our simulations lead to a 
similar dependence of the swimming speed on the number of colloids attached to the resin as seen in experiments, both regarding the qualitative 
shape of the curve and the saturation speed. 
The agreement is equally good for tilted substrates.
Still, some deviations are notable of course which is owed to the fact that the present model is a minimal model, 
attempting to distill the key ingredients underlying modular microswimming from a complex experiment, rather than describing all details 
of the experiment within a massively complex model; among other factors, the present model neglects to some extend the impact of the substrate wall
on hydrodynamic and phoretic fields, the extended size of resin and colloids, substrate friction and finally 
hydrodynamic interactions as induced by the gravitational force acting on resin and 
colloids (these are rather weak due to the relatively small tilt angles used). 

\section{Conclusions}
The present work provides a detailed picture
for the self-assembly and swimming mechanism of modular microswimmers
which has so far been illusive, even qualitatively. 
This picture involves both hydrodynamic and phoretic interactions between the involved modules (colloids, resin), 
which we have summarized in a 
simple model describing
the whole self-assembly and swimming behaviour in near 
quantitative agreement with our experiments. 
Our results may help 
describing other swimmers \cite{42}, may inspire new perspectives regarding 
the discussion of the relative strength of chemi- and electrophoretic 
effects in synthetic microswimmers \cite{43,44} and might be useful for
theories
predicting active self-assembly \cite{45,46,47} often involving phoretic 
interactions of unknown strength - 
the present work admits estimating them for modular microswimmers.

\begin{acknowledgments}
We thank Aidan T. Brown and Thomas Speck for useful discussions and Joost de Graaf for useful discussions and helpful comments on the manuscript. We acknowledge funding from DFG with the priority program SPP1726 (grant numbers LO418-17-2 and PA 459/18-2).
\end{acknowledgments}

%


\begin{thebibliography}{10}
\bibitem{38}%
  \BibitemOpen
  \bibinfo {journal} {Screening of this field would leave the system with a
  diffusion-induced charge imbalance}%
  \bibAnnoteFile{NoStop}{38}%
\end{thebibliography}

\end{document}